\documentclass[twocolumn,english,aps,prl,superscriptaddress,showpacs,preprintnumbers]{revtex4}
\pdfoutput=1
\usepackage[latin9]{inputenc}
\usepackage{units}
\usepackage{amsmath}
\usepackage{graphicx}
\usepackage{amssymb}

\makeatletter

\usepackage{nicefrac}

\makeatletter
\usepackage{babel}
\makeatother

\makeatother

\usepackage{babel}

\begin{document}

\title{Theoretical analysis of mechanical displacement measurement using a multiple cavity
mode transducer}

\author{J.M. Dobrindt}

\affiliation{Max Planck Institut für Quantenoptik, D-85748 Garching, Germany}

\author{T.J. Kippenberg}

\email{tobias.kippenberg@epfl.ch}

\affiliation{Max Planck Institut für Quantenoptik, D-85748 Garching, Germany}

\affiliation{Ecole Polytechnique Fédérale de Lausanne (EPFL), CH-1015 Lausanne,
Switzerland}
\begin{abstract}
We present an optomechanical displacement transducer, that relies
on three cavity modes parametrically coupled to a mechanical oscillator and whose
frequency spacing matches the mechanical resonance frequency. The additional resonances
allow to reach the standard quantum limit at substantially lower input power
(compared to the case of only one resonance), as both, sensitivity
and quantum backaction are enhanced. Furthermore, it is shown that 
in the case of multiple cavity modes, coupling between the modes is 
induced via reservoir interaction, e.g., enabling quantum backaction 
noise cancellation. Experimental implementation of the
schemes is discussed in both the optical and microwave domain.
\end{abstract}


\maketitle
%\title{Quantum limited detection of mechanical motion of mesoscale mechanical oscillators... quantum back-action and measurement of mechanical motion using a multiple cavity %mode transducer}

\begin{figure}[ptb]
\begin{centering}
\includegraphics[width=0.9\columnwidth]{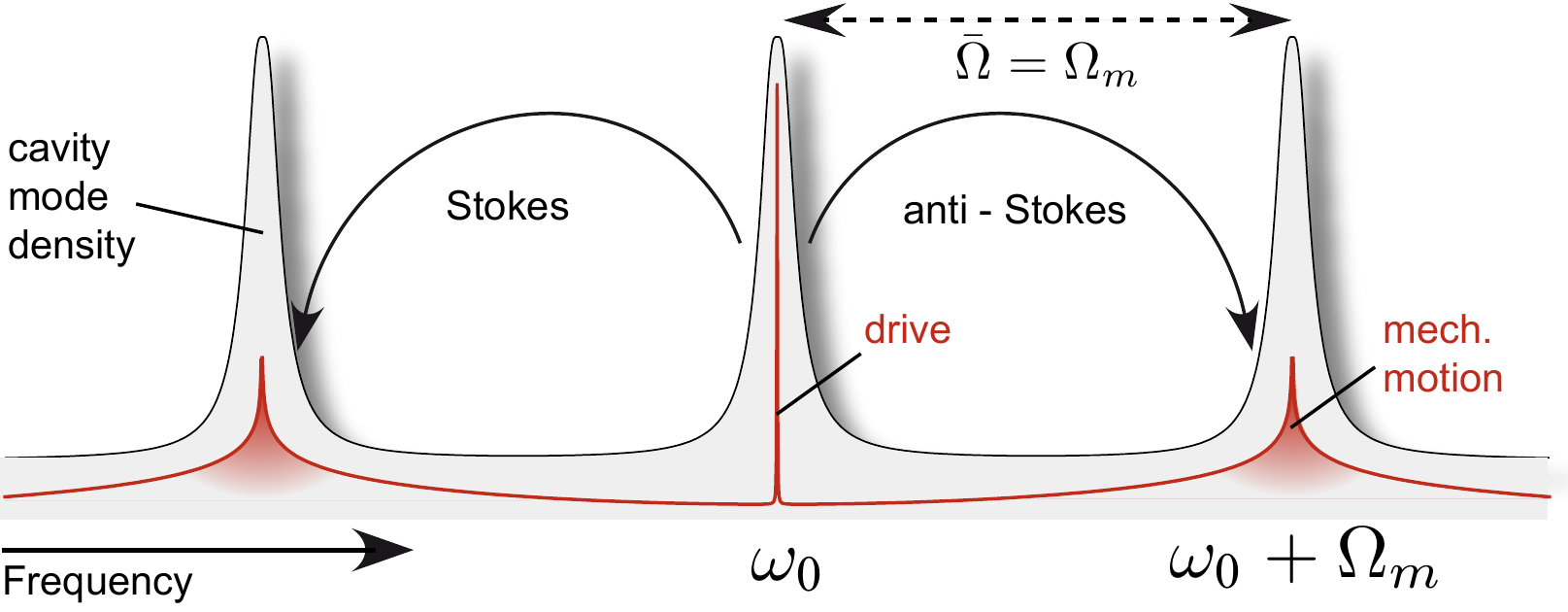} 
\par\end{centering}

\caption{Illustration of the triple mode transducer scheme. Auxiliary cavity
resonances at $\omega_{0}\pm\Omega_{\rm{m}}$ permit resonant motional side band
build-up. \label{fig:intro}}

\end{figure}

\textit{Introduction.}--- High frequency nano- and micro-mechanical
oscillators have received a high degree of attention recently. They
have been used as sensitive detectors, e.g. for spin \cite{rugar_single_2004}
or particle mass \cite{naik_towards_2009}, but also carry
an intrinsic interest in the study of small scale dissipation of mechanical
systems \cite{steele_strong_2009}, quantum limited motion detection \cite{teufel_nanomechanical_2009}, 
and backaction cooling of vibrational modes \cite{kippenberg_cavity_2008}. These studies have in common 
that a sensitive motion transduction is required, which can be implemented 
by parametric coupling to an optical, electrical, or microwave resonator. 
The ideal transducer should i) have a high
sensitivity and possibly operate at the standard quantum limit(SQL),
and ii) should operate at low power. The latter is experimentally advantageous,
as high power may cause excess heating due to intrinsic losses. The former
pertains to the minimum uncertainty in motion detection and arises
from the trade off between measurement imprecision, inherent to the
meter (i.e., detector shot noise), and (for linear continuous measurements)
inevitable \textit{quantum backaction} (QBA) \citep{caves_quantum_1982,clerk_introduction_2008}.
These processes are characterized by the displacement spectral density
$\bar{S}_{xx}(\Omega)$ and the QBA force spectral density $\bar{S}_{FF}(\Omega)$%
\footnote{The bar denotes a symmetrized spectral density: $\bar{S}(\Omega)=\nicefrac{1}{2}\left(S(\Omega)+S(-\Omega)\right)$.%
}. For a parametric motion transducer, where a single cavity mode
(with frequency $\omega_{0}$ and energy decay rate $\kappa$) is parametrically
coupled to a mechanical oscillator \citep{braginsky_measurement_1977},
the spectral densities are given by \begin{align}
\bar{S}_{xx}(\Omega) & =\frac{\kappa^{2}\hbar\omega_{0}}{64G^{2}P}\left(1+\frac{4\Omega^{2}}{\kappa^{2}}\right)\label{eq:Sx1}\\
\bar{S}_{FF}(\Omega) & =\frac{16\hbar G^{2}P}{\kappa^{2}\omega_{0}}\left(1+\frac{4\Omega^{2}}{\kappa^{2}}\right)^{-1}.\nonumber \end{align}
Here $P$ is the input power and the optomechanical coupling strength is determined by the cavity frequency
shift due to mechanical displacement: $G=\frac{d\omega_{0}}{dx}$. Equations \ref{eq:Sx1} satisfy 
$\sqrt{S_{xx}[\Omega]S_{FF}[\Omega]}\geq\hbar/2$, which is a consequence of  the Heisenberg uncertainty 
principle \citep{Braginsky_quantum_1992}.
The canonical way to lower the power to reach the SQL is to increase the cavity finesse, i.e., decreasing $\kappa$. 
However, Eqs. \ref{eq:Sx1} reveals a fundamental deficiency: decreasing
$\kappa$ for fixed $P$ only improves readout sensitivity as long as 
the mechanical signal (frequency $\Omega_{\rm{m}}$) lies within the cavity bandwidth, 
i.e., $\Omega_{\rm{m}}<\kappa$, while for $\Omega_{\rm{m}}>\kappa$ the displacement
sensitivity experiences saturation. Physically this phenomenon is
readily understood; the mechanical motion modulates the cavity field
and creates motional sidebands at $\omega_{0}\pm\Omega_{\rm{m}}$, which constitute
the readout signal. For $\kappa\ll\Omega_{\rm{m}}$, i.e., in the resolved
sideband regime (RSB), the sidebands (and therefore the signal) are
suppressed. This regime has recently been subject to experimental 
investigation \citep{schliesser_resolved-sideband_2008,teufel_nanomechanical_2009}.\\
% However this is only true as long as the mechanical signal
% falls within the cavity bandwidth, i.e. $\Omega_{\rm{m}}<\kappa$. Beyond
% this, in the so called resolved sideband regime (RSB) where $\Omega_{\rm{m}}>\kappa$,
% the mechanical signal is suppressed by the cavity cut-off. Due to
% the high frequency of micro oscillators, this regime has become experimentally
% accessible.\\
Here we present a readout scheme where this fundamental limitation is overcome,
by placing two auxiliary cavity resonances at $\omega_{0}\pm\Omega_{\rm{m}}$
around the central, driven resonance [cf. Fig. \ref{fig:intro}]. 
This enables resonant side band build-up and causes a substantial decrease in the power required
to reach the SQL. %
Moreover, we show that this scheme, when applied to the case of two
resonances, can lead to quantum backaction interference, without the
requirement of having a dissipative parametric coupling \cite{elste_quantum_2009}. 

\textit{Theoretical model.}--- In this section we present the theoretical
framework, to describe multiple cavity modes parametrically coupled to a mechanical
degree of freedom, described by its frequency $\Omega_{{\rm m}}$ and effective mass
$m_{{\rm eff}}$. This model covers a wide range of experimental
implementations on both nano- and microscale, as well as in the optical
and electrical domain. The cavity features several equidistant 
modes at frequencies $\omega_{k}=\omega_{0}+k\cdot\bar{\Omega}$  ($k\in\mathbb{Z}$),
described by the annihilation(creation) operators $\hat{a}_{k}$ $(\hat{a}_{k}^{\dagger})$,
where $\bar{\Omega}$ denotes the spacing between adjacent modes.
A driving field at frequency $\omega_{0}$ (input power $P$)
is coupled to the central cavity mode. Furthermore the optical modes are
parametrically coupled to the mechanical degree of freedom, $\hat{a}_{m}$
$(\hat{a}_{m}^{\dagger})$ and zero point motion $x_{0}=\sqrt{\hbar/2m_{{\rm eff}}\Omega_{{\rm m}}}$, via the interaction Hamiltonian \cite{Law1995}\begin{equation}
\hat{H}_{{\rm int}}=\hbar x_{0}\sum_{k,l}G\hat{a}_{k}^{\dagger}\hat{a}_{l}(\hat{a}_{m}+\hat{a}_{m}^{\dagger}).\label{eq:Hint}\end{equation}
The geometric factor, coming from the mode overlap integral, is assumed
(for simplicity) to be unity. Cavity
damping is modeled by coupling the cavity modes to a harmonic oscillator
bath via the damping Hamiltonian\begin{equation}
\hat{H}_{{\rm damp}}=i\hbar\sum_{k}\int_{-\infty}^{+\infty}d\omega\left[g_{k}^{\ast}(\omega)\hat{b}_{\omega}\hat{a}_{k}^{\dagger}-g_{k}(\omega)\hat{a}_{k}\hat{b}_{\omega}^{\dagger}\right].\label{eq:Hdamp}\end{equation}
The bath operators obey the commutation relations $[\hat{b}_{\omega},\hat{b}_{\omega'}^{\dagger}]=\delta(\omega-\omega')$.
In the following we will consider a classical harmonic oscillator
characterized by the position $\hat{q}=x_{0}(\hat{a}_{m}+\hat{a}_{m}^{\dagger})$
and damping rate $\gamma_{\rm{m}}$.
This treatment is justified, as we are solely interested in the transduction
properties of the cavity, and the quantum backaction coming from the
quantized nature of the field.\\
We eliminate the bath in the Markovian limit \cite{gardiner_quantum_2000}
setting $g_{k}(\omega)=\sqrt{\kappa}$. Surprisingly
the damping Hamiltonian does not only couple the cavity modes to the
dissipative bath, but also couples the modes among each other via
the reservoir dynamics {[}cf. Eqs. \ref{eq:HLE}{]}. This off-resonant
interaction is well known in laser theory (where it is responsible
for Petermann excess noise \cite{siegmann_excess_1989}), but has (to the authors knowledge) never
been applied to the context of opto- or electromechanics.\\
In the next step, we derive the Heisenberg-Langevin equations
(HLE) for the optical modes, where the classical drive is eliminated by moving to a rotating frame at the drive
frequency and subsequently transforming to the general quadrature
fluctuations $\hat{X}_{k,\theta}\equiv e^{-i\theta}(\hat{a}_{k}-\langle\hat{a}_{k}\rangle)+e^{i\theta}(\hat{a}_{k}^{\dagger}-\langle\hat{a}_{k}^{\dagger}\rangle)$.
We emphasize, that choosing one global rotating frame for all modes
is essential, as it enables us to treat off-resonant interaction terms
(to first order). These are known to account for quantum limits \cite{Dobrindt_parametric_2008}.
Explicitly the linearized HLE for the canonical quadrature fluctuations
$\hat{X}_{k}=\hat{X}_{k,\theta=0}$ and $\hat{Y}_{k}=\hat{X}_{k,\theta=\pi/2}$
are\begin{eqnarray}
\dot{\hat{X}}_{k} & = & -k\cdot\bar{\Omega}\hat{Y}_{k}-\frac{\kappa}{2}\sum_{l}\hat{X}_{l}+\sqrt{\kappa}\delta\hat{X}^{{\rm in}}[t]\nonumber \\
\dot{\hat{Y}}_{k} & = & k\cdot\bar{\Omega}\hat{X}_{k}-\frac{\kappa}{2}\sum_{l}\hat{Y}_{l}+g_{m}\hat{q}[t]/x_{0}+\sqrt{\kappa}\delta\hat{Y}^{{\rm in}}[t]\nonumber \\
\ddot{\hat{q}} & = & -\gamma_{\rm{m}}\dot{\hat{q}}-\Omega_{\rm{m}}^{2}\hat{q}+x_{0}g_{m}\sum_{l}\hat{X}_{l}.\label{eq:HLE}\end{eqnarray}
Solving for the canonical quadratures allows us to transform to the
$\theta$-dependent general quadrature. The global phase of the input
field is chosen in the way that $\bar{\alpha}=\Sigma\langle\hat{a}_{j}\rangle$
and the optomechanical coupling rate $g_{m}=2Gx_{0}\bar{\alpha}$
are real%
\footnote{This is always possible for a non-squeezed input field.%
}. The noise operators in the HLE are $\delta$-correlated: $\langle\delta\hat{X}^{{\rm in}}[t]\delta\hat{X}^{{\rm in}}[t^{\prime}]\rangle=\langle\delta\hat{Y}^{{\rm in}}[t]\delta\hat{Y}^{{\rm in}}[t^{\prime}]\rangle=\delta(t-t')$,
$\langle\delta\hat{X}^{{\rm in}}[t]\delta\hat{Y}^{{\rm in}}[t^{\prime}]\rangle=\langle\delta\hat{Y}^{{\rm in}}[t]\delta\hat{X}^{{\rm in}}[t^{\prime}]\rangle^{\ast}=\imath\delta(t-t')$.
We can account for intrinsic cavity loss by introducing a second loss
channel in Eqs. \ref{eq:HLE}, characterized by the internal loss
rate $\kappa_{0}$. It will appear in the results as the degree of
overcoupling $\eta_{c}=\kappa_{0}/\kappa_{{\rm tot}}$, with $\kappa_{{\rm tot}}$
being the total cavity decay rate%
\footnote{In the limit of overcoupling, $\eta_{c}\to1$.%
}. For multiple cavity modes, the output quadrature fluctuations are
given by a generalized input-output relation \cite{viviescas_field_2003}
\begin{equation}
\hat{X}_{\theta}^{{\rm out}}+\hat{X}_{\theta}^{{\rm in}}=\sqrt{\kappa}\sum_{k}\hat{X}_{k,\theta}\label{eq:multiIO}\end{equation}

\textit{Triple mode transducer.}--- Having introduced the theoretical
model, we calculate the output spectrum of a cavity with three optical
modes, spaced by the mechanical resonance frequency, i.e. $\bar{\Omega}=\Omega_{\rm{m}}$.
When the central resonance is driven, side bands at $\omega_{0}\pm\Omega_{\rm{m}}$,
that encode for the mechanical motion, build up efficiently {[}cf.
Fig. \ref{fig:intro}{]} and signal to noise is enhanced. \\
\begin{figure}[ptb]
\begin{centering}
\includegraphics[width=0.9\columnwidth]{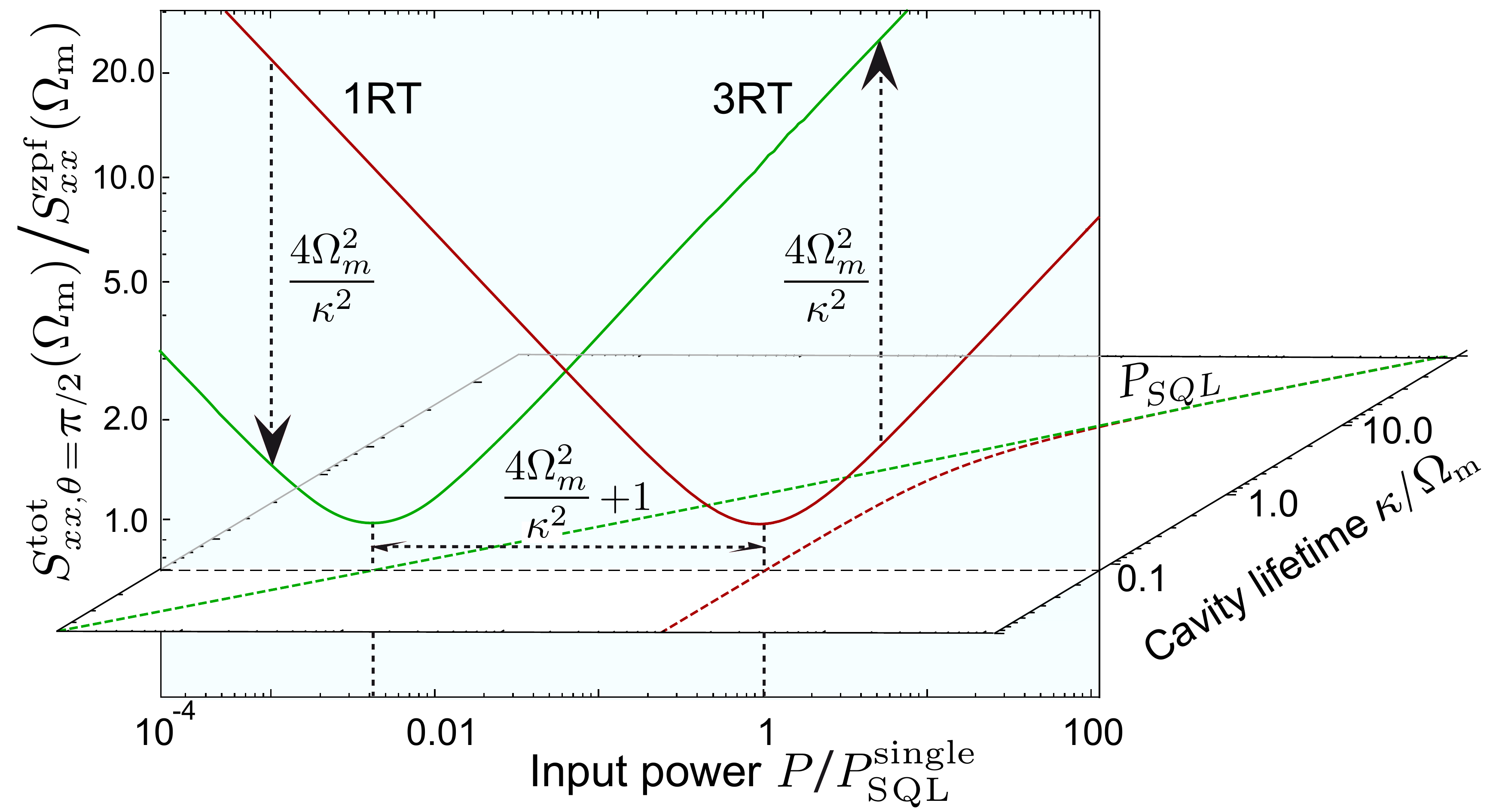} 
\par\end{centering}

\caption{The measurement noise spectrum $S_{xx,\theta=\pi/2}^{\rm{tot}}(\Omega_{\rm{m}})$
(normalized to $S_{xx}^{\rm{zpf}}(\Omega_{\rm{m}})=\hbar |\chi_{0}(\Omega_{\rm{m}})|$) for the triple cavity mode transducer as a function of the input power 
(with $\kappa=\Omega_{\rm{m}}/10$) compared to the single mode case. The traces in the horizontal plane
follow the minima as a function of the cavity life time $\kappa^{-1}$.
\label{fig:1}}

\end{figure}
From Eqs. \ref{eq:HLE} we calculate the quadrature fluctuations in
Fourier space. Using the multimode input-output relation (Eq. \ref{eq:multiIO}),
the spectrum of the output fluctuations is derived. Importantly the
off-resonant reservoir coupling terms ($\propto\kappa/2$ in Eqs.
\ref{eq:HLE}) preserve a flat shot noise spectrum for the decoupled
($g_{m}=0$) cavity. The measurement noise spectrum $S_{xx,\theta}^{\rm{tot}}(\Omega)=S_{xx,\theta}(\Omega)+\left|\chi_{0}(\Omega)\right|^{2}S_{FF}(\Omega)$
is obtained by scaling the output fluctuations to the mechanical signal \cite{jaekel_quantum_1990}.
The classical motion of the mechanical oscillator, characterized by
its bare susceptibility $\chi_{0}(\Omega)$, is not affected by the
coupling, as dynamical backaction effects are absent \cite{kells_considerations_2002}.
The shot noise background is given by \begin{equation}
S_{xx,\theta}^{\rm{triple}}(\Omega)=\frac{1}{\sin^{2}\theta}\frac{x_{0}^{2}\kappa}{\eta_{c}g_{m}^{2}}\left(1+\frac{4\Omega^{2}\left(\Omega_{\rm{m}}^{2}-\Omega^{2}\right)^{2}}{\kappa^{2}\left(\Omega_{\rm{m}}^{2}-3\Omega^{2}\right)^{2}}\right).\label{eq:lambda3RT}\end{equation}
The sensitivity is maximized for $\theta =\pi/2$, implying
that the information on the mechanical signal is encoded in the phase
quadrature. Comparing $S_{xx,\theta}^{\rm{triple}}(\Omega)$ to the single
resonance transducer, we note that the transduction properties of a 
low frequency signal remain unchanged, i.e., $S_{xx,\theta}^{\rm{single}}(0)/S_{xx,\theta}^{\rm{triple}}(0)=1$.
However the sensitivity at the mechanical resonance frequency is dramatically
increased\begin{equation}
S_{xx,\theta}^{\rm{single}}(\Omega_{\rm{m}})/S_{xx,\theta}^{\rm{triple}}(\Omega_{\rm{m}})=1+\frac{4\Omega_{\rm{m}}^{2}}{\kappa^{2}}.\end{equation}
For systems that operate well into the RSB
regime, such as toroidal microresonators \cite{schliesser_resolved-sideband_2008}
or superconducting microwave resonators \cite{teufel_nanomechanical_2009}, 
this factor is more than $\times100$ and thus represents
a major reduction. Based on the Heisenberg uncertainty principle for
continuous position measurements, one expects that the enhanced sensitivity
also implies an increased quantum backaction force spectral density.
Physically these can be viewed as the beat of the carrier (at $\omega_{0}$) with vacuum
fluctuations at $\omega_{0}+\Omega_{\rm{m}}$, which resonantly heat the
mechanical oscillator. The radiation pressure force fluctuations are given by\begin{equation}
\delta\hat{F}^{{\rm rp}}[\Omega]=\frac{\hbar g_{m}}{2x_{0}}\sum_{k}\hat{X}_{k}[\Omega].\label{eq:radPress}\end{equation}
Indeed, in the same way, that the shot noise is reduced, the backaction force
spectral density $S_{FF}^{\rm{triple}}(\Omega)$ is increased, and the
Heisenberg limit of the single resonance transducer is recovered.
\begin{equation}
\sqrt{\bar{S}_{xx,\theta=\pi/2}^{{\rm triple}}(\Omega)\cdot\bar{S}_{FF}^{{\rm triple}}(\Omega)}=\frac{\hbar}{2\sqrt{\eta_{{\rm c}}}}.\label{eq:SQL3RT}\end{equation}
We emphasize that the recovery of the Heisenberg
limit is a consequence of the off-resonant coupling. Moreover, the triple transducer
has the significant advantage over a single cavity mode transducer
that the SQL is reached at substantially lower power: $P_{{\rm SQL}}^{{\rm single}}/P_{{\rm SQL}}^{{\rm triple}}\approx4\Omega_{\rm{m}}^{2}/\kappa^{2}$ 
\footnote{We note, that the mixing term $S_{xF,\theta}^{{\rm triple}}(\Omega)$
is odd in $\Omega$ for $\theta=\pi/2$ and therefore disappears in
the symmetrized spectrum. %
}. Moreover, the enhanced QBA itself can be a valuable resource. Indeed,
many quantum optomechanical experiments rely on QBA to be the dominant
force noise, such as in experiments relating to ponderomotive squeezing \cite{verlot_probing_2008}
or two beam entanglement \cite{mancini_entangling_2002}.

\textit{Dual mode scheme.}--- Within the framework of the multimode
transducer theory, we can also consider the situation of two resonances,
spaced by the mechanical resonance frequency \cite{braginsky_parametric_2001,matsko_sensitivity_2008}.
The situation differs from the triple mode scheme, as only the anti-Stokes
process is resonantly enhanced by pumping the lower frequency mode.
This results in net cooling of the mechanical degree of freedom, because
every scattering process annihilates one phonon%
\footnote{On the other hand, pumping the higher frequency mode results in amplification.%
}. Then the response to an external force, e.g. a thermal Langevin
force, a signal force, or quantum Langevin forces, is suppressed as a result of the damped mechanical motion.
Consequently the transduction properties are not ideal. \\
Next, we calculate the QBA spectral density from Eq. \ref{eq:radPress}.
\begin{eqnarray}
S_{FF}^{{\rm dual}}(\Omega) & = & \frac{\hbar^{2}}{x_{0}^{2}}\frac{g_{m}^{2}\kappa(\Omega_{\rm{m}}-2\Omega)^{2}}{4(\Omega_{\rm{m}}-\Omega)^{2}\Omega^{2}+\kappa^{2}(\Omega_{\rm{m}}-2\Omega)^{2}}\end{eqnarray}
\begin{figure}
\begin{centering}
\includegraphics[width=0.9\columnwidth]{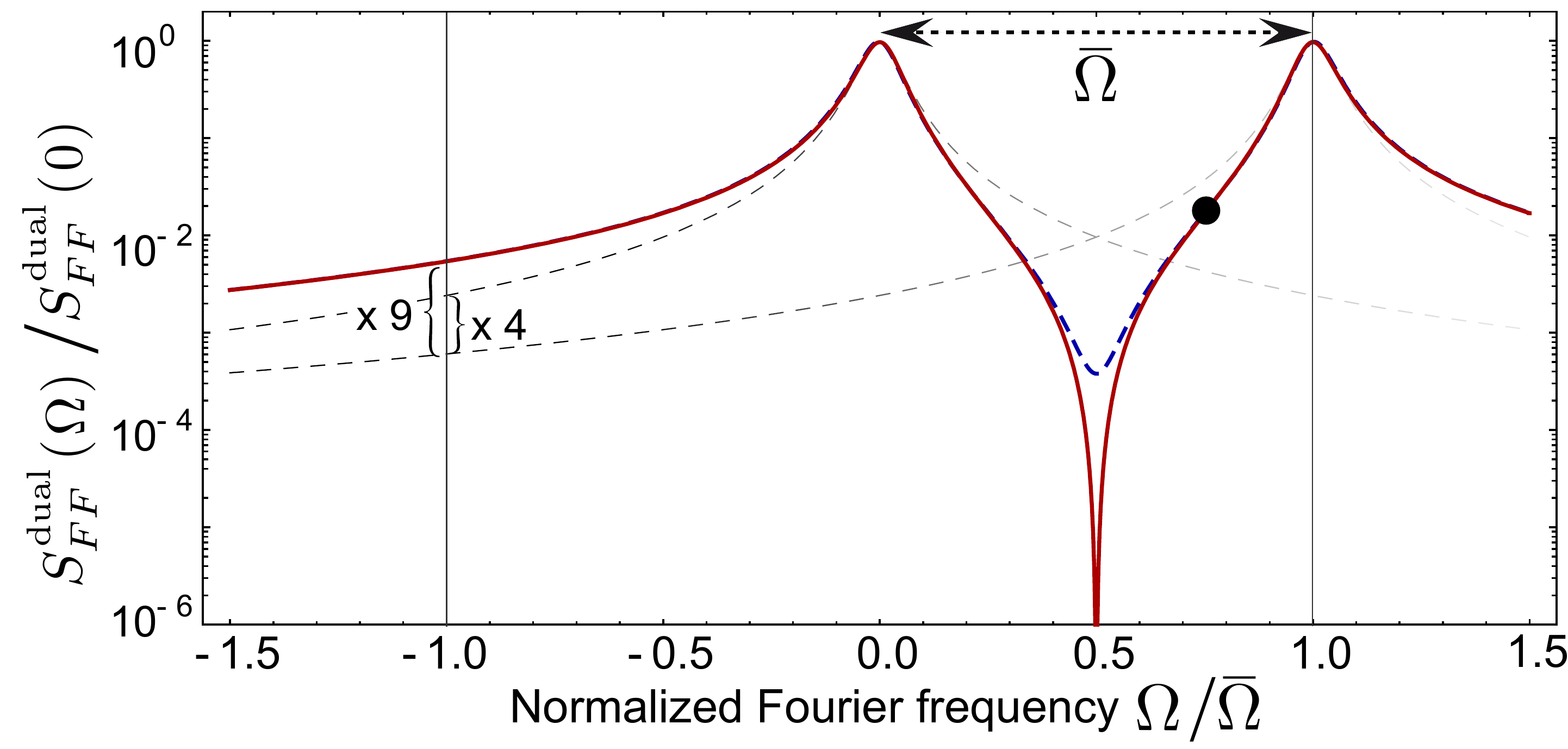} 
\par\end{centering}
\caption{The normalized quantum backaction force spectral density $S_{FF}^{{\rm dual}}(\Omega)/S_{FF}^{{\rm dual}}(+\Omega_{\rm{m}})$
(thick solid line) is plotted together with the result from a classical model (thick dashed line).
The reservoir interaction between the modes leads to complete noise
cancellation at $\Omega=\Omega_{\rm{m}}/2$. The thin dashed lines indicate
the backaction force coming from two uncorrelated modes. \label{fig:2}}

\end{figure}
Then an estimate for the final occupancy of the mechanical mode $n_{f}=\left\langle \hat{a}_{m}^{\dagger}\hat{a}_{m}\right\rangle $
is given by the quantum noise approach \cite{clerk_introduction_2008}.
Unexpectedly, compared to the single resonance dynamical backaction
cooling \citep{marquardt_quantum_2007,WilsonRae_theory_2007}, the
quantum limit increases by a factor of $\times9$.\begin{equation}
\frac{n_{f}}{n_{f}+1}=\frac{S_{FF}^{\rm{dual}}(-\Omega_{\rm{m}})}{S_{FF}^{\rm{dual}}(+\Omega_{\rm{m}})},\Rightarrow n_{f}\approx9\,\frac{\kappa^{2}}{16\Omega_{\rm{m}}^{2}}\label{eq:coolLimit}\end{equation}
This is understood from the constructive quantum noise interference
at $\Omega=-\Omega_{\rm{m}}$ {[}cf. Fig \ref{fig:2}, $(\sqrt{1}+\sqrt{4})^{2}=9${]}.
However, the QBA spectrum in Fig. \ref{fig:2} reveals an additional,
striking feature. At $\Omega=\Omega_{\rm{m}}/2$ the \textit{quantum noise
exactly cancels}. This is a direct consequence of the reservoir coupling
terms in Eqs. \ref{eq:HLE}. Omitting these terms yields a \textit{classical}
interference pattern, shown by the thick dashed curve in Fig. \ref{fig:2}.
The shape of the QBA spectrum suggests to tune the mode spacing
to $\bar{\Omega}=4\Omega_{\rm{m}}$ and drive the cavity on the red wing
of the upper resonance {[}cf. Fig. \ref{fig:2}, black circle{]}.
Then the heating term vanishes, i.e., $S_{FF}^{\rm{dual}}(-\Omega_{\rm{m}})=0$
(in the rotating frame). However, an exact analysis, using a covariance
approach \cite{wilson-rae_cavity_2008}, reveals that finite line width
effects lead to a power dependent quantum limit, i.e. $n_{f}\approx2g_{m}^{2}/\Omega_{\rm{m}}^{2}$.
As the cooling rate saturates, when $g_{m}$ approaches $\kappa/2$,
one can find an upper limit for $n_{f}$ by assuming $g_{m}<\kappa/2$ \cite{Dobrindt_parametric_2008}.
The same analysis for the canonical two mode cooling yields $n_{f}\approx(9\kappa^{2}+14g_{m}^{2})/16\Omega_{\rm{m}}^{2}$.\\
With respect to the transduction properties of the dual scheme
we note, that in cooling experiments the QBA is not viewed as additional
measurement noise (as for the 3RT), but contributes the signal itself.
This is corroborated in the case of ground state cooling, where the
field fluctuations conserve the zero point fluctuations of the mechanical
oscillator, independently of its quantum nature. 

\textit{Experimental implementation.}--- The experimental challenge
in the design of a multimode transducer lies in matching the cavity
mode spacing with the resonance frequency of the mechanical oscillator without adding additional
damping. The canonical setup is a Fabry-Pérot cavity where the free
spectral range matches the resonance frequency of the harmonically
suspended back mirror. However difficulties might arise from differing
mode overlap integrals. These challenges can be circumvented in a more general
way, adaptable to optical, electrical, and microwave domain.%
\begin{figure}
\begin{centering}
\includegraphics[width=0.9\columnwidth]{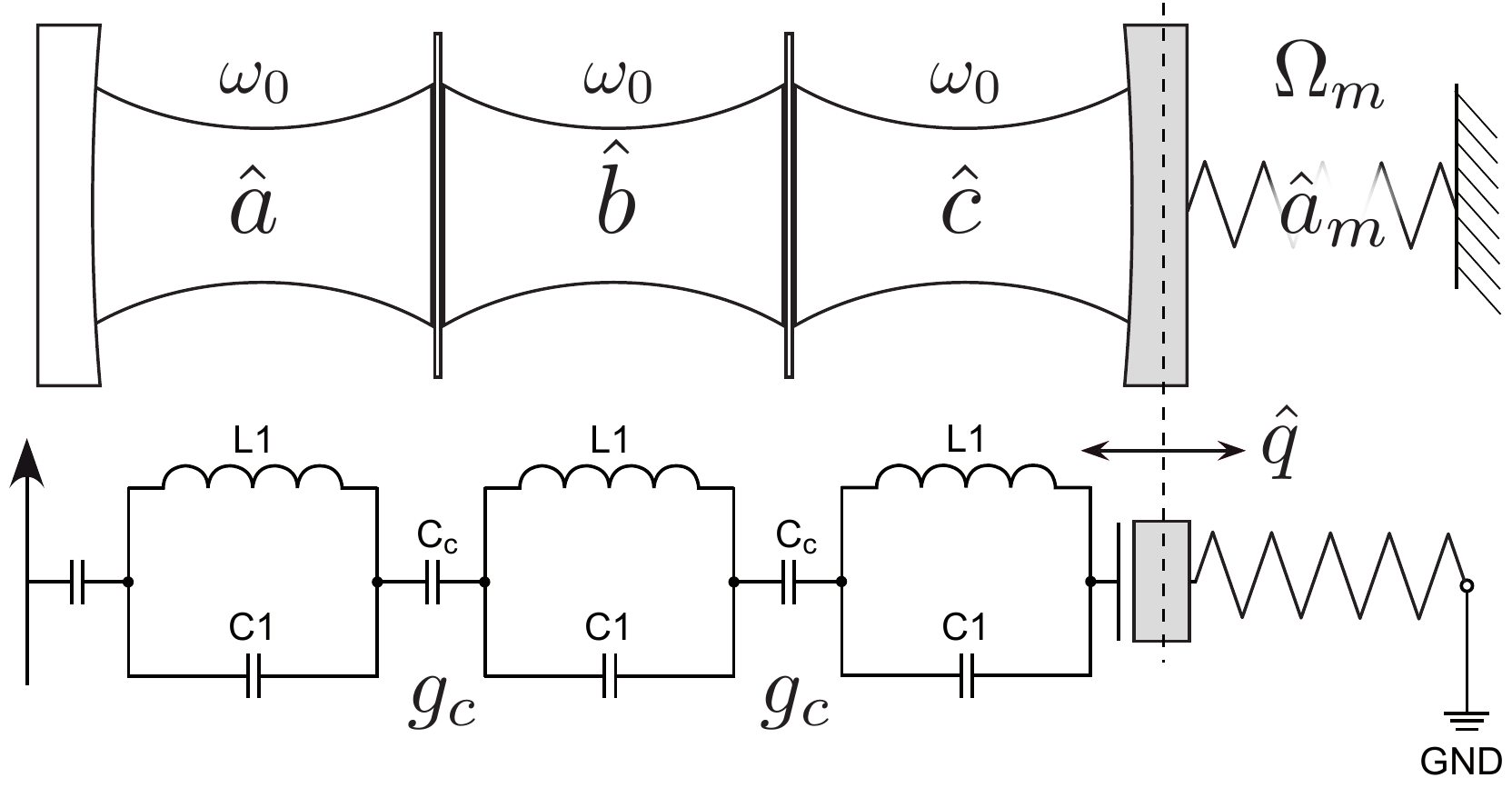} 
\par\end{centering}

\caption{a) Three degenerate optical modes are coupled via semi-transparent
mirrors. b) Three LC-oscillators (possibly microwave resonators) are
capacitively coupled {}``in series''. While mode $\hat{a}$ is connected
to an external drive, i.e., a transmission line, mode $\hat{c}$ is
parametrically coupled to a mechanical oscillator. When the coupling
rate $g_{c}>\kappa$ exceeds the individual decay rate, the spectrum
exhibits normal mode splitting. \label{fig:3}}

\end{figure}
As illustrated in Fig. \ref{fig:3}, three degenerate cavity modes $\{\hat{a},\hat{b},\hat{c}\}$
are coupled \textit{in series} via the linear interaction $\hbar g_{c}\left[\hat{b}(\hat{a}^{\dagger}+\hat{c}^{\dagger})+\hat{b}^{\dagger}(\hat{a}+\hat{c})\right]$.
In the microwave domain, this interaction can be realized by coupling
of three superconducting quarter or half wave resonators (via inductive
or capacitive coupling as shown in Fig. \ref{fig:3}). In the optical
domain, it can be achieved by coupling of degenerate cavity modes via partially transparent mirrors or evanescent field.
In addition, only one mode ($\hat{c}$) is coupled to the mechanics
by $\hat{H}_{{\rm int}}^{{\rm single}}=\hbar x_{0}G\,\hat{c}^{\dagger}\hat{c}(\hat{a}_{m}+\hat{a}_{m}^{\dagger})$.
In the regime of strong mode coupling, when $g_{c}>\kappa$ , the
originally degenerate cavity modes exhibit normal mode splitting. The new cavity
eigenmodes can then be represented in a basis of dressed states $\left\{ \hat{a}_{0},\hat{a}_{+},\hat{a}_{-}\right\} $
with eigenfrequencies $\left\{ \omega_{0},\omega_{0}\pm\sqrt{2}g_{c}\right\} $.\begin{equation}
\left(\begin{array}{c}
\hat{a}_{0}\\
\hat{a}_{-}\\
\hat{a}_{+}\end{array}\right)\propto\left(\begin{array}{ccc}
0 & -1 & 1\\
-\nicefrac{1}{\sqrt{2}} & \nicefrac{1}{2} & \nicefrac{1}{2}\\
\nicefrac{1}{\sqrt{2}} & \nicefrac{1}{2} & \nicefrac{1}{2}\end{array}\right)\cdot\left(\begin{array}{c}
\hat{a}\\
\hat{b}\\
\hat{c}\end{array}\right).\label{eq:dressedStates}\end{equation}
The splitting $\sqrt{2}g_{c}$ can be matched to the mechanical resonance
frequency by appropriately tuning the coupling rate. Transforming to the
dressed state basis, one finds that the operator $(\hat{c}\propto\hat{a}_{0}+\hat{a}_{-}+\hat{a}_{+})$
is proportional to the sum of the dressed state operators {[}cf. inverse
of matrix in Eq. \ref{eq:dressedStates}{]}. Replacing $\hat{c}$ in
the parametric interaction $\hat{H}_{{\rm int}}^{{\rm single}}$ results
in a multimode interaction as given by Eq. \ref{eq:Hint}.
Indeed, a dual mode coupling of this kind has recently been demonstrated
using toroidal microcavities \cite{grudinin_phonon_2009} in this proposed way. Moreover,
tunable mode splitting between counter propagating modes has also
been achieved \cite{anetsberger_near-field_2009}, making the experimental realization of this new class of high frequency transducers realistic.\\
We acknowledge Warwick Bowen for discussion and making helpful and critical comments.

\bibliographystyle{apsrev}

\end{document}